\documentclass[twocolumn,floatfix,reprint]{revtex4}
\usepackage{graphicx,amssymb,amsmath,placeins}

\begin{document}

\title{A dominance tree approach to systems of cities}

\author{Thomas Louail}
\email{thomas.louail@cnrs.fr}
\affiliation{CNRS, G\'eographie-cit\'es, Campus Condorcet, 93322 Aubervilliers Cedex, France}

\author{Marc Barthelemy}
\email{marc.barthelemy@ipht.fr}
\affiliation{Institut de Physique Th\'{e}orique, CEA, CNRS-URA 2306, F-91191, 
Gif-sur-Yvette, France}
\affiliation{CAMS (CNRS/EHESS) 54 Avenue de Raspail, 75006 Paris, France}

\begin{abstract}

  Characterizing the spatial organization of urban systems is a challenge which
  points to the more general problem of describing marked point processes in
  spatial statistics. We propose a non-parametric method that goes beyond
  standard tools of point pattern analysis and which is based on a mapping
  between the points and a `dominance tree', constructed from a recursive
  analysis of their Voronoi tessellation. Using toy models, we show that the
  height of a node in this tree encodes both its mark and the structure of its
  neighborhood, reflecting its importance in the system. We use historical
  population data in France (1876-2018) and the US (1880-2010) and show that the
  method highlights multiscale urban dynamics experienced by these
  countries. These include non-monotonous city trajectories in the US, as
  revealed by the evolution of their height in the tree. We show that the height
  of a city in the tree is less sensitive to different statistical definitions
  of cities than its rank in the urban hierarchy. The method also captures the
  attraction basins of cities at successive scales, and while in both countries
  these basin sizes become more homogeneous at larger scales, they are also more
  heterogeneous in France than in the US.  Finally, we introduce a simple
  graphical representation – the height clock – that monitors the evolution of
  the role of each city in its country.
\end{abstract}

\maketitle


Understanding the organization of urban systems has always been a central
challenge in geography and economics
\cite{Helsley:1990,Sanders:1997,Rozenblat:2007}. On one hand the population
sizes of cities have been extensively discussed since Zipf's work
\cite{Zipf:1949,Gabaix:1999,Verbavatz:2020}, and have been shown to follow a
broad distribution: there is a hierarchy of cities characterized by many small
cities, a few medium cities and a very few large ones, whose population sizes
are much larger than the rest. While this distribution has been fully
characterized and discussed \cite{Cottineau:2017,Verbavatz:2020}, the spatial
distribution of cities has also been an important subject of debate
\cite{Christaller:1933,Ullman:1941,Haggett:1965,Gonzalez-Val:2019}. 

More generally, this problem of characterizing the spatial organization of urban
systems appears often in spatial statistics \cite{Illian:2008,Jacobsen:2006}
where the points (cities in our case) have a position and are described by (at
least) one quantity such as the population. In spatial statistics this general
setting is referred to as a `marked point process'~\cite{Illian:2008}, where
each point $x_i$ of the process carries extra information called a mark
$m_i$. The mark can be a random or a categorical variable, or any other
additional information about the points. This problem is not only important in
geography with the study of human settlements, but is also relevant for many
fields ranging from ecology (positions of plants of different species), to
epidemiology (locations of infected individuals), material science (positions of
defects) or astronomy that is interested in the location of stars and
galaxies. An urban system is an example of such marked point process, where the
points are the cities and the mark is their population size. Space is obviously
important in this system, and has to be considered jointly with population: if
for example two cities are close to each other, it makes a difference if they
have similar or very different population sizes. Standard tools developed for
the analysis of point processes usually consist in measuring spatial
autocorrelations \cite{Moran:1950}, or testing the null hypothesis of complete
spatial randomness with the $K$ or $L$ statistics which summarize the deviations
from a uniform (Poisson) distribution \cite{Illian:2008,Jacobsen:2006}. These
tools were then extended to marked point processes and describe deviations from
known cases such as the Poisson process, or intensity and moments measures
\cite{Haenggi:2012}. A tool that would go beyond these measures and provide a
more precise characterization of these processes would then be extremely useful
for a wealth of different problems.

To further characterize marked point processes it is possible to start from
geometrical structures constructed on top of the point pattern. These `secondary
structures' \cite{Illian:2008} comprise in particular tessellations and networks
\cite{Okabe:2000}. The Voronoi tessellation is one of the most relevant
structure in computational geometry, and is of major importance in the
resolution of many problems, in particular in location science
\cite{Laporte:book}. Networks can also be constructed over a set of points, and
useful tools include proximity graphs (such as the random geometric graph) or
excluded volume graphs (such as the Gabriel graph). Measures on these secondary
structures can then characterize the point process itself, and constructing a
spatial network on top of the point process enables to import all the networks
knowledge into spatial statistics. For example, a recent approach uses
first-passage times of random walks on networks constructed over a set of points
in order to quantify correlations in complex systems \cite{Bassolas:2020}.
However, approaches based on secondary structures are in general used to
estimate deviations from uniformity, or the importance of some correlations. In
contrast, we want to construct a tool that helps us understand the multiscale
structure of a system of marked points, and enable us to compare it to other
systems. Such a tool for characterizing the organization of spatial hierarchies
should thus encode both the spatial information and the population (in more
general terms both the positions and the marks of the points). The purpose of
this tool is not to describe the deviation from a uniform distribution, but to
compare two systems (such as urban systems in different countries for example)
and to understand the temporal evolution of these systems from a non-local,
high-level perspective. We will illustrate the construction of such a tool by
considering the case of urban systems.

Cities are not scattered at random but follow some logic based on geographical
constraints, economical considerations and historical path dependency, and
central place theory \cite{Christaller:1933} has seeked to explain the spatial
distribution of cities of different sizes based on the idea that settlements
function as `central places' that provide services to surrounding areas. The
result of consumers' preferences is then a system of centers of various sizes,
forming different levels of a hierarchy. A consequence obtained by Christaller
\cite{Christaller:1933} is that the most efficient pattern to serve areas
without overlap is a triangular or hexagonal arrangement of
settlements. Although this idea has been very successful and inspirational to
many scientists, few works have tried to validate it quantitatively. For
example, cities have been studied from the correlation point of view
\cite{Glass:1971,Hernando:2014}, but an important contribution to this problem
is due to Okabe and Sadahiro \cite{Okabe:1996} who showed that random (uniform)
arrangements of cities could explain Christaller's findings. To reach this
result they had to define a quantitative tool that captures the spatial
dominance of a city on another. They used a tree representation of the dominance
relation between marked points, and they used it to characterize both a Poisson
point pattern and a real-world case study --- market places in Nishinomiya, near
Osaka. They analyzed this tree in the perspective of testing Christaller's ideas
and in particular measured the ratio $K$ between the number of nodes in two
consecutive hierarchical levels. However they did not go further in the
characterization of the importance of each point in the system, and in the
following we will adapt and extend this method in order to construct a general
tool able to characterize a system of marked points. First we will describe the
dominance tree introduced in \cite{Okabe:1996}, and that will constitute the
starting point of our analysis. We will then introduce different measures to
extract information from this tree. In particular, we will show that the height
of a node in this tree encodes both its mark and space-related information. For
systems of cities, this means that the height of city in the dominance tree
encodes both its population rank and its location in space. We will first
illustrate this method on toy models, and then on empirical data, to compare the
evolution of the French and the US urban systems over the 20th century.

\section*{Constructing the dominance tree}

The method we propose can be applied to any marked point process,
and we will illustrate it on the case of cities in a country. We assume that a
given country has $N$ cities and each city $i$ is characterized by its location
$x_i$ in some coordinate system, and its population $P_i(t)$ which can vary over
time. In order to characterize this system we have to define a data structure
that encodes both the spatial information (the location of cities) and their
importance (their population size).

\begin{figure*}[htb]
\centering
      \includegraphics[width=0.9\textwidth]{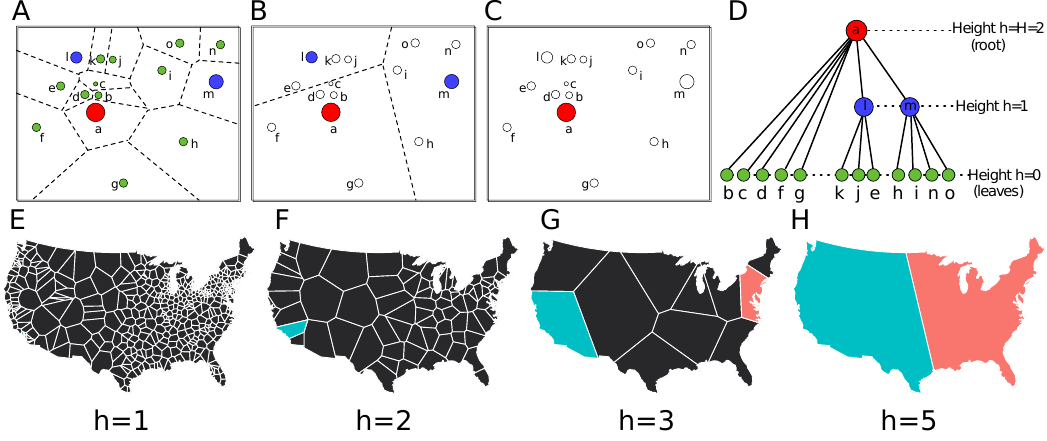}
      \caption{Construction of the dominance tree. (A) Each point is
        characterized by a mark/value represented here as the size of the
        circle. We first construct the Voronoi tessellation (shown with dotted
        lines) for these points and we observe that there are three local
        maxima: $a$, $l$, $m$, in red and blue. These are the points which do
        not have any neighbor whose mark is larger than their own. (B) From step
        (A) we keep the local maxima only, we construct the Voronoi tessellation
        over this set, and determine the local maxima (which is here the node
        $a$). (C) At the end of the process, we are left with the node with the
        largest mark. (D) The entire process can be described by the `dominance
        tree': leaves correspond to the points which are not local maxima, and
        the children of a given node are the points that belong to its
        attraction basin. For example, as can be seen in (B), nodes $k$, $j$ and
        $e$ belong to the attraction basin of node $l$, which is a local maximum
        at level/height h=1. (E-H) Successive Voronoi tesselations obtained when
        applying this `decimation' process for US cities marked with their 2010
        population sizes~\cite{USdata} (we didn't show the case $h=0$ which
        corresponds to the Voronoi tessellation of all US cities and which is
        unreadable). After $H=6$ iterations, we obtain a single city. By
        construction, the node with the largest mark will always be the root of
        the dominance tree, but for the other nodes their height in this tree
        will not only depend on their mark but also on their location.}
\label{fig:voroexample}
\end{figure*}

The first step is to construct the {\it dominance tree} proposed by Okabe and
Sadahiro \cite{Okabe:1996}. The idea is to recursively construct Voronoi
tesselations over the set of nodes (see for example the book \cite{Okabe:2000}
and references therein). The Voronoi cell $V_i$ of a node $i$ is the set of
points that are closer to $i$ than to any other node,
$V_i=\{\;x\;|\;d(x,x_i)<d(x,x_j), \forall j\neq i\}$. We show in
Fig.~\ref{fig:voroexample}(A) an example of a Voronoi tesselation computed for
nodes distributed in the plane. Starting from this Voronoi tessellation, we
identify local maxima or local `centers': a point $i$ is a local maximum if its
population is larger than those of the neighboring Voronoi cells. In other
words, for any point $i$ we define with the help of the Voronoi tesselation the
set of neighbors $\Gamma(i)$ that are the nodes whose Voronoi cell is adjacent
to $V_i$. A city $i$ is then a local center if its population $P_i$ is larger
than the populations of its neighbors: $P_i>P_j\;\;\forall\;j\in\Gamma(i)$. In
the example of Fig.~\ref{fig:voroexample}(A), we thus have $3$ local maxima
labeled $a$, $l$, and $m$. In a second step, we keep these local centers and
construct a new Voronoi tessellation over them
(Fig.~\ref{fig:voroexample}(B)). Cities that do not appear anymore (i.e. that
were not local centers at the previous step) belong now in the new Voronoi cell
of a local center, and are therefore `dominated' by this city and belong to its
`attraction basin' (for example, nodes $b$, $c$, $d$, $f$, $g$, belong to the
attraction basin of node $a$).  Here again we determine local maxima, i.e. cells
that do not have any neighbor whose population is larger than their own. We
repeat this procedure recursively until only one city is left
(Fig.~\ref{fig:voroexample}(C)) which, by construction, is the city with the
largest population. We therefore understand that a city will `survive' many
iterations if its population is large but also if it is well located. A large
city located very close to an even larger one will quickly be absorbed by this
larger city. It is interesting to note that this process bear some ressemblance
to the coarse-graining obtained by means of real-space renormalization
\cite{Kadanoff:2000}. The dominance tree -- shown in
Fig.~\ref{fig:voroexample}(D) for the simple case we just described -- is then
some sort of bookkeeping of the changes of scale and the hierarchical
organization of the marked points: each node has children that correspond to
other nodes which belong to its attraction basin. We show on
Fig.~\ref{fig:voroexample}(E-H) the Voronoi tesselations that correspond to the
successive steps of this process applied to US cities. In this case the root is
New York City (successive Voronoi cells of New York and Los Angeles are colored)
and the depth of the tree, that is the number of iterations to reach the root,
is $H=6$.

We thus represent this process by the dominance tree
(Fig.~\ref{fig:voroexample}(D)) where we keep at each level the remaining points
and where the links denote the spatial dominance relation. At height zero (the
leaves of the tree), we then have all cities of the system that are locally
dominated by a larger neighbor city. After one iteration of the process, we have
the local centers, etc. until we reach the root of the tree which is the city
with the largest population. Okabe and Sadahiro proposed this construction
\cite{Okabe:1996} and used it to prove that the most important quantitative
statements in Christaller's central place theory were also observed for a random
spatial Poisson point process. More precisely, for both the random Poisson point
patterns and their real world case study – marketplaces in Nishinomiya, Japan –
they measured Christaller’s K indicator, which is the ratio between the number
of local centers at consecutive hierarchical levels. They also calculated the
statistics of the number of edges of the Voronoi cells as a function of the
number of points in the point pattern, and for different families of point
patterns. Consequently their characterization of the structure of the dominance
tree was limited, and mainly focused on geometrical aspects.

. 

We can then represent a system of cities by a dominance tree, where the
height of a city is the largest iteration before it is absorbed by a larger
city. Once we have constructed the dominance tree, each city $i$ has its height
$h_i$. We recall here that, by definition, the height of a node in a tree is the
number of edges between this node and the furthest leaf going down in the
tree. In other words, it is the length of the longest path (i.e. its number of
edges) from the node to the deepest leaf (in contrast, the depth of a node in a
tree is the distance from a node to the root). The height of a leaf is then
$h=0$, while the root has the largest height (denoted by $H$ in the
following). We will show that, due to the statistical properties of Voronoi
tesselations built from spatial Poisson point processes, $H\approx \log_6(N)$
where $N$ is the number of points in the system, and H is in general of order $5$
or $6$ in our empirical analysis of urban systems. We introduce
$n_f(h)$ which is the number of cities such that their final height is $h$. In
contrast, the total number of cities at a given height $h$ is denoted by
$n_i(h)$ and we have the following relation $n_i(h)=\sum_{h'\geq h}n_f(h')$.

Once we have constructed the dominance tree, we have to characterize it. There
is a large number of possible measures, but we will focus here on the height of
a city, and we will discuss it for toy models and for empirical data. A city $i$
at time $t$ is then characterized by its population $P_i(t)$ (or equivalently by
its rank $r_i(t)$ when population are sorted by decreasing order), and its
height $h_i(t)$ in the dominance tree. This height characterizes the role of
the city in the hierarchical organization of the urban system. While in the
following we analyze urban systems at the national scale, the method could be
applied at different scales as well, such as the regional or continental
scale.

\section*{Toy models}

In order to define a toy model we have to specify both the point distribution
and the population distribution. Different models are possible and we will
mainly consider the following variants. First, we will consider that populations
are distributed according to the power law $\rho(P)\sim 1/P^\alpha$ with
$\alpha=2$. This case corresponds to the classical result of Zipf and even if
recent data show that this exponent can fluctuate considerably
\cite{Cottineau:2017,Verbavatz:2020}, it won't affect our discussion here, as
the relevant quantity is not the population size itself, but its corresponding
rank in the hierarchy: indeed, what matters for the construction of the
dominance tree is to know if a city is larger than its neighbors. In the
following, we will thus indifferently discuss population or rank for
characterizing a given city.

For the point distribution we will consider a uniform distribution (spatial
Poisson process) of the $x_i$ in the square $[-1,+1]^2$. We thus have two lists:
a list of population sizes $P_i$ and a list of positions $x_i$
($i=1,\dots,N$). In order to define a model we have to specify how to match
these lists, and the resulting correlations encode in a simple way the spatial
organization of the system. 
To simplify the description we will assume that
the populations are sorted in decreasing order $P_1>P_2>\dots >P_N$ and that
locations are sorted in increasing order according to their distance to the
center $(0,0)$: $||x_1||< ||x_2||<\dots <||x_N||$. We will consider the
following three cases:
\begin{itemize}
\item The `deterministic model' : we associate the largest population to the closest
  point to the center $(0,0)$: $P_1\leftrightarrow x_1$ and follow the order
  $P_i\leftrightarrow x_i$. This model mimics in some way a system with a
  central organization around the main city.
\item The `random model': we associate the largest population to the closest
  point to the center $(0,0)$: $P_1\leftrightarrow x_1$, and then associate
  randomly the rest of the $P_i$s to the rest of $x_i$s. In this case, we have a
  central large city but there is no specific organization around it.
\item Finally, the `tunable model' is less deterministic. We first assign $P_1$
  to $x_1$ as before, but for the remaining $N-1$ cities we proceed as
  follows. For $P_2$ we choose at random a position among
  $\{x_2,x_3,\dots,x_N\}$ according to the probability
  $p_k=\exp(-d(x_1,x_k)/L)/Z$ where $d(x_1,x_k)$ is the distance between $x_1$
  and $x_k$, $Z=\sum_{k=2,...,n} \exp(-d(x_1,x_k)/L)$ is the normalization and
  $L$ is a positive parameter that can be interpreted as a polarization or
  interaction distance. We then proceed in a similar way for $P_3$ with the
  remaining $x_i$ left, etc. until the list of populations is exhausted. This
  tunable model is then able to interpolate between the deterministic case with
  $L\ll 1$ and the random case with $L\gg 1$.
\end{itemize}

\begin{figure}[!ht]
  \centering
  \includegraphics[width=\linewidth]{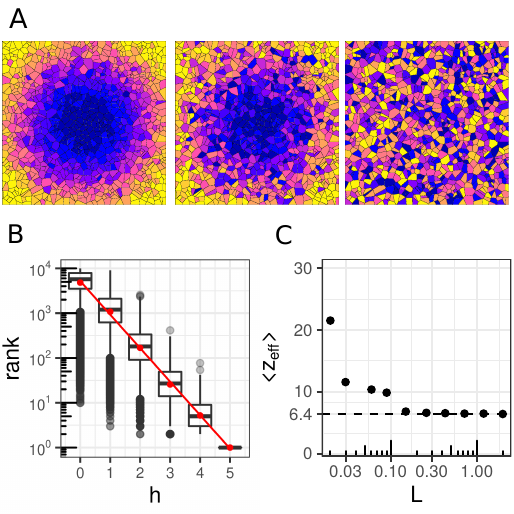}
  \caption{(A) Initial Voronoi tessellations computed for $N=10^3$ points. From
    left to right, the tunable toy model interpolates from the deterministic
    model (the population decreases with the distance to the central largest
    city) to the random model where distance and population to the center are
    uncorrelated. (B) Rank versus height for the random toy model, computed for
    $N=10^4$ cities and 200 disorder configurations. The straight line is a
    linear fit of the form $\langle\log (r)\rangle = 8.555 - 1.722*h$
    ($R^2>0.99$). The points show the outliers and the lower and upper sides of
    the boxes correspond to the second quartile and third quartile respectively
    (the boxes therefore contain half of the points). The middle bar corresponds
    to the median. (C) Effective connectivity of the Voronoi tessellation for
    different values of $L$ in the tunable model.}
\label{fig:2}
\end{figure}

These toy models allow us to explore the effect of various parameters and to
test various aspects of our characterization of spatial hierarchies. We show in
Fig.~\ref{fig:2}(A) the Voronoi tessellation of $10^3$ points in the square. The
color code represents here the rank of cities according to their population (the
darker, the larger the associated population). On Fig.~\ref{fig:2} we have
represented the `tunable' model that interpolates from the deterministic case
(on the left) with a clear `monocentric' pattern with population decreasing with
the distance to the center, to the random model (last figure on the right) where
there are no correlations between rank and distance to the center.

Before turning to these specific cases, we first discuss some general properties
about the depth of the tree, and the number of nodes at a certain level.  First,
by construction, the root of the dominance tree is the city with the largest
population. An important quantity is the depth $H$ of the tree that will measure
the number of different hierarchical levels in the system. In order to estimate
$H$, we first evaluate the number of cities $n_i(h)$ that are left at the height
$h$. We denote by $z_h$ the average number of neighbors in the Voronoi
tessellation at level $h$, and the number of cities at level $h$ is then
\begin{align}
  n_i(h)&=\frac{N}{z_1z_2\dots z_h}\simeq N\mathrm{e}^{-h\langle\ln z\rangle}
\end{align}
where the brackets $\langle\cdot\rangle$ denote the average over all
levels. This expression implies that the number of cities and the level $h$ are
related as follows
\begin{align}
 \ln n_i=a-bh
  \label{eq:hvsn}
\end{align}
where $a=\ln N$ and $b=\langle\ln z\rangle\equiv \ln z_{\mathrm{eff}}$.  The
depth of the tree $H$ corresponds to the height of its root, at which there is
only city left ($n_i=1$). Hence we obtain 
\begin{align}
  H=\frac{\ln N}{\langle\ln z\rangle}
\end{align}
In the `random' toy model, cities are uniformly distributed and we can roughly
expect $z_{\mathrm{eff}}\approx 6$~\cite{Okabe:2000} and with $N=10^4$ cities we
obtain a tree of height $H\approx 5$. In contrast, for the deterministic model,
there is essentially a single local maximum leading to a small depth (see
Supplementary Figures S2 and S3)

We consider now the relation between the rank and the height of a city in the
dominance tree, and the result is shown for the random model in
Fig.~\ref{fig:2}(B) where locations are fixed and different disorder
realizations correspond to different arrangement of populations sizes on these
fixed locations (in other words, location is the quenched disorder, while the
mark is the annealed disorder). This plot shows that on average, smaller ranks
lead indeed to larger heights (and eventually the largest city with rank $r=1$
has the largest height $h=H$). However, we also observe on this plot that this
relation is not univocal, as there are large fluctuations around the average
behavior. In other words, the rank (or population) of a city does not determine
its height in the tree. This shows that the height encodes simultaneously the
rank of a city in the urban hierarchy and its location.  In particular, we see
on Fig.~\ref{fig:2}(B) that cities with ranks $r\geq 2$ can have different
heights depending on the disorder realization.

For each value of $L$ for the tunable model, using the general relation
Eq.~\ref{eq:hvsn}, from the measure of $n_i(h)$ versus $h$, we compute the
average number $z_{\mathrm{eff}}=\exp(b)$ of neighbors of the Voronoi
tessellations obtained in the decimation process (see Fig.~\ref{fig:2}(C)). For
small $L$ values, this model reproduces the deterministic case where cities are
distributed around the largest one in decreasing order. In this case in many
regions there are no local maxima and the effective number of neighbors
$z_{\mathrm{eff}}$ is very large. As $L$ increases we tend towards the random
model where cities are distributed randomly in the plane. In this
case 
we expect a regular Voronoi tessellation constructed over a Poisson point
process, with $z_{\mathrm{eff}}\approx 6$ \cite{Okabe:2000}.

\begin{figure}[ht!]
  \centering
  \includegraphics[width=\linewidth]{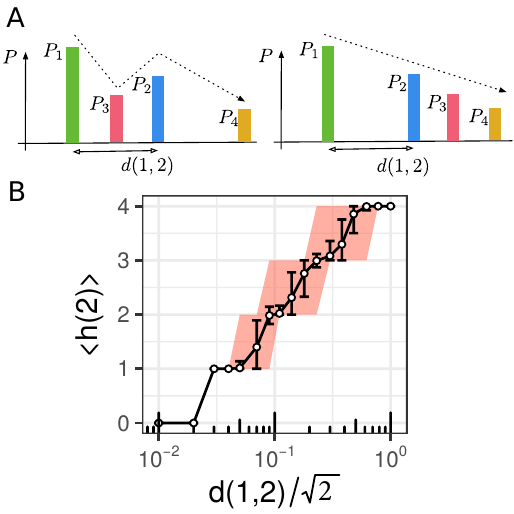}
  \caption{(A) Two different possible configurations: (left) smaller cities are
    located between cities $1$ and $2$ or (right) no cities are in between. In
    the former case, the city $2$ is a local maximum, while in the latter, city
    $2$ belongs to the dominance basin of city $1$. (B) Average value of the
    height $h_2$ of city $2$ as a function of its distance $d(1,2)$ to the
    largest city ($P_1$) in the random model, normalized by the largest distance
    possible. Error bars represent here $\max(h_2)-\min(h_2)$ observed over
    $200$ random configurations for the cities. }
\label{fig:h2}
\end{figure}

We have shown on these simple toy models that the height of a city in the tree
does not depend on its rank alone, but also on its location and its
neighborhood. In order to illustrate further the effect of the spatial
arrangement of cities on their heights, we consider the following simple
situation. We locate the largest city $1$ (of rank $r=1$) at the center of the
plane and we assign to a point at distance $d(1,2)$ the second largest city $2$.
We then locate at random all the other cities and measure the height $h_2$ of
city $2$ as a function of the distance $d(1,2)$. We also record the maximum and
minimum height reached by $2$ in the random configurations. The resulting plot
is shown in Fig.~\ref{fig:h2}.
%
We see in this plot that the height $h(2)$ of city $2$ depends strongly on the
spatial arrangement of cities in the space between it and the larger city ($1$):
if there is a smaller city in between, the city $2$ is a local maximum and has a
larger height (Fig.~\ref{fig:h2}(A)). In the opposite case, city $2$ will be
absorbed by city $1$ at the next iteration of the decimation process. We test
this numerically and the result is shown in Fig.~\ref{fig:h2}(B). First, as
expected we observe that the height $h(2)$ is on average an increasing function
of the distance to the largest city. At small distance, the city $2$ has a large
chance to belong to the attraction basin of city $1$ and will be quickly
`absorbed' by it, leading to a small height. In contrast, as the distance
increases it will take more steps before the city $2$ is in the attraction basin
of city $1$, leading to a larger height. In addition, for a given distance
$d(1,2)$ we observe important fluctuations of $h(2)$. This clearly shows that
the height encodes the spatial organization of the neighborhood of a city, and
in particular of the space between the city and the closest larger one. The
height is then related to the number of local maxima in this intermediate space.

These toy models demonstrate the relevance of the height of a city in the
dominant tree for monitoring its importance in the system. We will now apply
these ideas on empirical data. 

\section*{Empirical studies: US and French urban systems}

The previous section helped us to understand the main properties of the
dominance tree and the factors that determine the height of a marked point. We
will now use this tool to measure the spatial organization of the French and the US
urban systems, and their evolution over the last 130 years (see the Data section
at the end for details).

\subsection*{Height fluctuations}
\begin{figure*}[ht!]
  \centering
  \includegraphics[width=\textwidth]{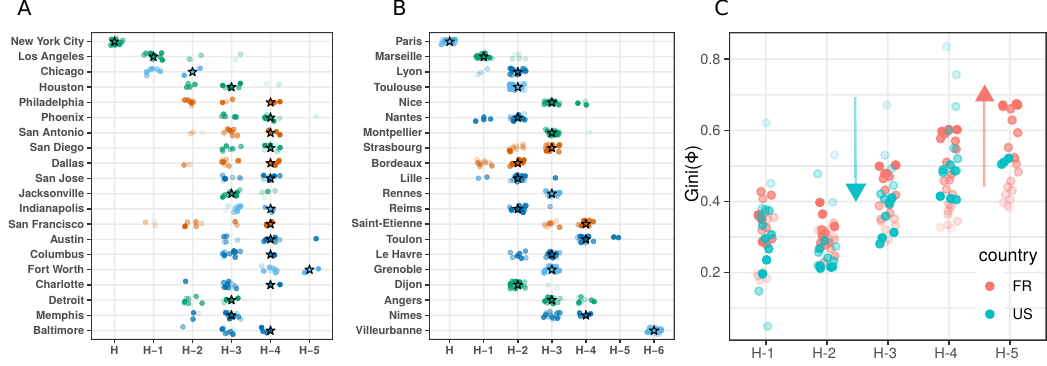}
  \caption{Height in the dominance tree for different cities in the US (A) and
    in France (B). The amount of transparency corresponds to the time period
    (following the chronological order from faded to clear) and the star denotes
    the city's height for the most recent year in our dataset (2010 for the US
    and 2018 for France). The color of the dots correspond to the class of
    cities they belong to, defined in Fig. \ref{fig:clock}. (C) Evolution of the
    Gini coefficient for the population sizes $\phi$ of the attraction basins of
    cities with a given height $h$. The results are shown for France and the US,
    and the opacity follows the chronological order (almost transparent points
    correspond to the earliest time periods).}
  \label{fig:rang_gini}
\end{figure*}

We first focus on the height of cities as discussed above. We show on
Fig.~\ref{fig:rang_gini}(A) and (B) the heights, at different dates, of the $20$
cities that are currently the largest ones in the US and in France.

The root of the tree is by construction the largest city of the system, which in
both countries has remained the same over the entire period --- New York City in
the US, and Paris in France. We observe small height variations for French large
cities, a fact that was already observed in \cite{Batty:2006} for the ranks of
old European cities in the urban hierarchy, but with notable exceptions such as
Lille. 
In the US it is interesting to note that even if the population of San Francisco
grew steadily since its beginning, its height in the US tree has decreased,
signalling a changing environment and the higher growth of neighboring cities.

We observe that in both countries there are different cities which at some point
of their history reached the height $H-1$ ($H$ is the root): in the US Los
Angeles, Chicago, or San Francisco; and for France, Marseille, Lyon, Nantes,
Bordeaux or Lille.  We also observe that the 20 municipalities that are
currently the largest ones in France have experienced much less height
fluctuations than the 20 largest US cities, some of the latter spanning 3 or 4
different levels in the tree. The height thus captures ascending and descending
city trajectories in the US that were not observed in France during the last 130
years. Like other countries in Europe, France has experienced a long-standing
urbanization, with a slow and regular evolution, and a city like Paris was
already an international hub in the Middle-Age. In contrast, the US is a
relatively new country, which experienced successive waves of urban development
in the last centuries \cite{Bretagnolle:2009}. The trajectory of a given city in
the dominance tree highlights how the role and the local importance of this city
has evolved in time. It provides a richer information that the rank alone
\cite{Batty:2006}, because it encapsulates information not only about the
position of the city in the urban hierarchy, but also about its importance among
its neighbors.

\medskip

\subsection*{Basin sizes and statistical definition of cities} 

\subsubsection*{Basin sizes: Gini and Zipf}
Beyond single cities trajectories, the dominance tree also contains information
about the spatial organization of the urban hierarchy at larger scales. At each
step of the tree construction process, 
a local maximum -- say city $i$ -- contains a number of cities in its Voronoi
cell $V(i)$.  We denote by $\phi_i$ the total population of this attraction
basin: $\phi_i=\sum_{j\in V(i)}P_j$ (the population of the `seed' city $P_i$ is
not taken into account in $\phi_i$).  At each step of the iterative construction
process we thus have a collection of values of $\phi$, one for each Voronoi
cell. We compute the Gini coefficient $G$ of these values (see for example
\cite{Gini}): a large value of $G$ (i.e. close to one) indicates that there are
a few very large basins of attraction and all the others are small, while in
contrast a small $G$ (close to zero) indicates that most attraction basins have
roughly the same population size. On Fig.~\ref{fig:rang_gini}(C) we plot $G$ for
France and the US at different tree levels $h$, that correspond to different
spatial scales, and for different time periods. There are two remarkable
features on this plot. First, in both countries the Gini is decreasing as we
approach the root, which means that even if we start from a very heterogeneous
situation at small scales (at the urban agglomeration level), at a large scale
the population sizes of the attraction basins become comparable.  Second, we
observe an important difference between the two countries. In France the system
has evolved towards a situation where the attraction basins are becoming more
heterogeneous (at all spatial scales), while the data show the opposite in the
US. Surprisingly enough, it seems that in France, the urban system has not
evolved towards a uniform distribution of important basins, but that inequalities
in population sizes have increased in the recent periods, at all spatial scales.
Furthermore the fact that Zipf's plots of $\phi$ for different levels (see
Supplementary Figure S1 and the corresponding section in the SI) are all well
described by power laws, confirms that the hierarchy of population sizes is
preserved at different scales. The power law fits give exponent values
displaying a decreasing trend, indicating less fluctuations when $h$ increases,
in agreement with the results obtained with the Gini coefficient.

\subsubsection*{Height sensitivity to different statistical definitions of cities}
\label{sec:sensitivity}

Defining consistent city boundaries is a central issue in urban studies,
especially when the goal is to characterize and compare
cities~\cite{Cottineau:2016,Cottineau:2017,Pumain:2015}. In general,
administrative boundaries such as municipalities fail to capture meaningful
borders, and many statistical definitions of cities have been proposed. Some are
based on morphological aspects (built-up area), some other on functional ones
(journey-to-work commuting flows, see Supplementary text in SI for details).
While different criteria lead to different boundaries, key variables that
quantify the importance of a city in the system should not depend on small
fluctuations at finer scales.  For example, for cities that are regional,
national or international hubs, key variables such as their (population) rank in
the urban hierarchy or their height in the dominance tree should not vary too
much for different city definitions. We will test this robustness for the french
case. In 2021, there were about $2,400$ \emph{urban units} (morphological
definition) in Metropolitan France, that gathered approx. $7,500$ municipalities
representing a total population of $50$M people. Following the functional
definition, there were $682$ \emph{urban areas}, gathering $26,000$
municipalities for a total population of $60$M people. For each definition we
determine the rank of cities in the population hierarchy, and then attribute
their rank to all the municipalities that compose the city according to this
definition. Since Paris' urban unit is the most populated in France, all the
municipalities that are part of Paris urban unit are attributed rank $1$, all
the municipalities included in Lyon's urban unit are attributed rank $2$,
etc. We do the same for the urban area definition: we determine the population
hierarchy of urban areas, and attribute their rank to all the municipalities
that compose the urban area. The 7k+ municipalities that are simultaneously part
of an urban area and of an urban unit thus have two population ranks $r_i$ and
$r'_i$, one according to the morphological definition, the other according to
the functional definition. In order to compare these lists of ranks, we compute
the Kendall correlation coefficient $\tau$ (see for example
\cite{Abdi:2007}). This quantity $\tau$ is equal to one when the order of the
cities is strictly identical in the two lists, and $-1$ when the lists have an
opposite order. Here, we obtain for these lists a value $\tau_r=0.28$. We then
compute the dominance tree for France according to the two different city
definitions, and determine the height of each city in both trees (we allocate
this height to all the municipalities that compose the city according to this
definition). For the 7k+ municipalities that are part of a urban unit and also
part of a urban area, we end up with two lists of heights, and we compute the
Kendall $\tau$ coefficient between these two lists. We obtain $\tau_h=0.30$, and
we thus have $\tau_h>\tau_r$, which shows that the height of a city in the
dominance tree is in fact slightly less sensitive to different city definitions
than its rank in the population hierarchy.

\smallskip

\subsection*{Relation between city size and height in the tree}
We also investigate the relation between the population size of a city and its
height in the dominance tree. In order to quantity this relation, we compute two
lists in which cities are ranked in decreasing order of (i) population sizes and
(ii) heights. We compare these lists using Kendall's $\tau$ correlation
coefficient. For both countries we calculate this
quantity for each year available in the data (see Supplementary Figure S4), and
we also calculate it for the toy models previously discussed. In particular, we
obtain for the `random' toy model the average value
$\langle \tau\rangle \approx 0.43$ (represented on Supplementary Figure S4 by
the dashed line).  For the determinisc, monocentric toy model we obtain a value
close to 0, and for the tunable model, $\tau$ interpolates between these two
values for different $L$.  Overall for the US and France, we observe relatively
small values of Kendall's $\tau$ in these systems, confirming that the height is
not determined by population alone, and that the location of a city has its own
importance and effect. Supplementary Figures S5 and S6 complement this
observation, and show that there are many pairs of cities such as $P_i > P_j$
and $h_i < h_j$.

We can understand these different results with the following
considerations. When large cities are close to one another, some are dominated
by even larger ones, and consequently their height in the tree is small. This
leads to a small value of $\tau$ as observed for the deterministic model. In
contrast, when the population size and the location are uncorrelated as in the
random model, large cities can reach higher levels in the tree before being
`absorbed' by an even larger city. In this case, there is a higher correlation
between rank and height, leading to a larger value of $\tau$. We observe that
$\tau$ is smaller for France than for the US, suggesting that the French system
of cities displays a stronger concentration of population within large urban
agglomerations. These are composed of large municipalities located close to one
another, in the vicinity of a historical large municipality that is the core of
the agglomeration. In addition, we observe that $\tau$ is decreasing in time for
both countries, signalling a decrease of the correlation between the height and
the rank in the urban hierarchy. This can be due to two different effects:
smaller cities becoming more important (large $h$) due to their strategical
location, or in contrast cities with small $h$ whose population size increased
because they are located near an important city from which they depend.

\medskip

\subsection*{Height clocks} A city $i$ of population $P_i(t)$ has thus a certain
height $h_i(t)$ in the dominance tree which characterizes the importance of its
role at a regional level. In contrast, the leaves of the tree (with height
$h=0$) do not dominate any other city and depend directly on a more important
city in their vicinity. 
In order to characterize the evolution of the urban system and how different
cities can see their role evolving in this system, we study the temporal
variation of their height. A first approach was shown in
Fig.~\ref{fig:rang_gini}(A,B), but for a better visualization we propose a
`height clock', in the same spirit as the rank clock introduced by Batty in
\cite{Batty:2006}. On a polar plot, the radius is equal to the normalized height
$max(h) - h_i(t)$ and the angle is proportional to time. During a time range
$[t_1,t_2]$, the total height variation for a city $i$ is then given by
$\Delta_i=\sum_{t=t_1}^{t_2-1} |h_i(t+1)-h_i(t)|$ and according to the dynamics
of this variation, we classify cities into four categories and show some
representative examples in Fig.~\ref{fig:clock} (see Supplementary Figure S7 for
the rank clocks of some selected cities).

\begin{figure}[ht!]
  \centering
\includegraphics[width=\linewidth]{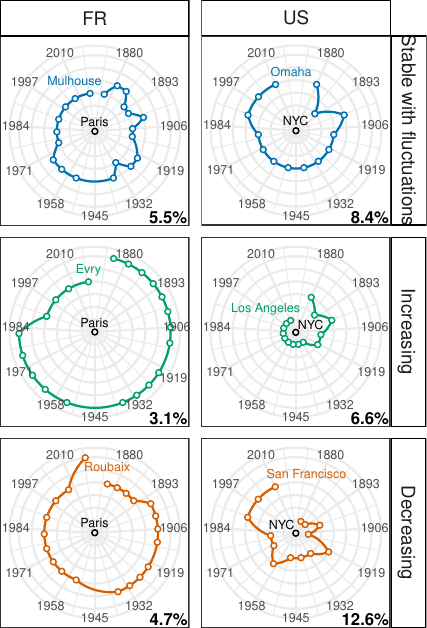}
\caption{Height clocks for French (left column) and US municipalities (right
  column). We illustrate here different classes of possible dynamics with
  selected cases: stable with fluctuations (top line), increasing height
  (middle) and decreasing height (bottom) in the dominance tree. On each plot,
  we indicate the percentage of cities in the urban system that belong to that
  class. The central node corresponds to the root of the system, i.e. the city
  with the largest height. The color code corresponds to the one used in
  Fig.~\ref{fig:rang_gini}(A) and (B).}
\label{fig:clock}
\end{figure}

The first type of cities display a constant height (`stable' cities, not shown
here). For France this is for example the case of Paris (and New York City for
the US), which was always the most important city in the system since
1880. 
Toulouse in France has a constant height equal to $4$ ($H-2$, which can also be
seen in Fig.~\ref{fig:rang_gini}), which indicates the steady importance of this
city at a regional level. This class of cities is by far the most important for
both countries with a share of $86.7\%$ for France and $72.4\%$ for the US. This
is an important indication that the French system reached earlier some kind of
more stable state, compared to the US where more active change dynamics were
observed in the last century.

The second type comprises cities for which $h(t_1)=h(t_2)$, but which
experienced height fluctuations in between. This is the case for example of
Mulhouse, Nantes or Lyon in France, or Omaha in the US. As we can see, Mulhouse
saw its height fluctuating between 1870 and 1970 before stabilizing in the last
decades. During the whole period its population grew steadily (except during the
2nd world war) and this again shows that the height analysis encodes more than
just the population dynamics, 
by considering also the population dynamics of neighboring cities. The third
type correspond to cities whose height decreased, while the fourth type are
cities whose height increased and which became more important in the
organization of the system. The proportion of cities that belong to one of these
two categories is larger in the US than in France, which is consistent with the
urban dynamics observed in both countries during the 20th
century~\cite{Bretagnolle:2009}. Obviously these results need to be replaced in
a geographical context, but these observations prove that non-trivial
information can be extracted through this single quantity $h$, that can shed
light on the evolution of cities and their importance in the urban system.

To summarize, the evolution in time of the height of a point in the dominance
tree sheds a light on the evolution of its importance in the spatial
hierarchy. This information is complementary to the time evolution of its rank
according to the value of its mark, as illustrated by
Supplementary Figure S8 that shows the height clock of three large
US cities alongside their rank clock~\cite{Batty:2006}. The height and the rank
trajectories of a given city can even go in opposite directions, as this is the
case for example for Portland and Dallas. 

\section*{Discussion}

An important problem in spatial statistics is to characterize a system of points
that have marks. For cities, the simplest mark is the population size and the
importance of a city at different geographical scales results from both its
location and its rank in the urban hierarchy, determined by its population
size. We presented a simple tool based on the dominance tree that encodes the
spatial structure of a marked point process. 


Starting from an original method proposed by \cite{Okabe:1996}, we have
introduced new metrics, toy models, and new vizualisations for this problem. In
particular, we have conducted a proper analysis of the influence of the spatial
distribution of marks in space on the maximal height a point with a given rank
can reach in the tree. We have introduced $\phi$, the population size of the
attraction basins, and we have also introduced the height clock that allows to
monitor the time evolution of the height of a particular city. Finally we have
also introduced the height vs. rank, and height vs. population size diagrams
(see Supplementary Figures S5 and S6).

The height of a point in this tree appears to be a crucial information, and its
monitoring allows to understand the dynamics of spatial hierarchies. We have
applied this method to toy models and to empirical analysis of the evolution of
the French and the US systems of cities. This method however goes well beyond
these examples and could be applied to a wealth of problems, depending on the
nature of points and on the quantity chosen to characterize them. It allows for
the study of spatial correlations in a simple way, and could help in comparing
the dynamics of many different urban systems, discussing their evolution in
time, classifying them, etc. and finally contributing quantitatively to urban
theory thanks to a shared, non-parametric tool. It could also be useful in
connecting distinct bodies of literature, namely the studies of the statistical
distribution of settlement and city sizes~\cite{Cottineau:2017, Verbavatz:2020},
and those that investigate their spatial distribution on the
Earth~\cite{Gonzalez-Val:2019, Strano:2020}. Further studies could also
investigate the relation between the height of cities and socio-economical
factors, such as their GDP.

Concerning the application of this method to urban systems, several lines of
future research can be identified. The first one would be to adapt the method so
that it could be applied not only to points but also to cellular tissues, by
considering the real geographical envelopes of the geographical units rather
than the Voronoi cells computed from their centroids. Since the set of
municipalities, departments, regions, etc. in a country naturally form
tessellations, the rest of the procedure would remain the same: identify the
cells that are the local maxima, aggregate the dominated cells to the one that
is a local maxima, and so on. At a smaller, intra-urban scale, it could also be applied to
capture the dynamical organisation of hotspots within the city.

Future studies may also further investigate the sensitivity of the method
against different definitions of cities, that would correspond to different
marked point processes or to different elementary building blocks used for
building the dominance
tree. 
In the same vein, it would be interesting to measure differences in the tree
structure when considering more elaborated distances than the Euclidean
distance. One could for example consider transport networks distances, or
average travel times. Such data are however much harder to obtain for many
different countries, and for ancient periods of time. Another direction could be
to analyze the sensitivity of scaling laws exponent for different hierarchical
levels in the dominance tree \cite{Xiao:2022}.

Beyond urban geography, other problems in spatial statistics could also be
addressed with this simple tool. As it encodes in a simple way various
correlations, it could shed a new light on the characterization of random
tessellations and random patterns in two dimensions, and could help in
revisiting important problems in the physics of foams, such as the Aboav-Weaire
and Lewis laws \cite{Chiu:1995,Weaire:1984} which in general focus on the
distribution of local statistics such as the number of edges, the neighbor
statistics, etc. but not on the global spatial organization.

\section*{Data}

We use census population data for France provided
by the French national statistics bureau (INSEE)~\cite{FRdata}. These
data give the population sizes of approximately $32,000$ French
municipalities (administrative boundaries) of Metropolitan France for
every decade during the period 1876-2018. For the US, the dataset
\cite{USdata} is a compilation of US cities populations between 1790
and 2010 (every 10 years). The data come primarily from the US Census
Bureau dataset and contain $7,500$ incorporated cities which at a
certain point of their history had a population larger than $2,500$
individuals. Other cities were added from a variety of sources (see
\cite{USdata} for details). Both datasets are public and can be freely 
downloaded  (see \cite{FRdata,USdata} for the URLs).

We note that the underlying criteria used to delineate cities in space
for computing their population size may influence our results. Here we use
administrative boundaries (municipalities) for France which are known to be
problematic in many ways when comparing historical population data from one city
to another, or between countries~\cite{Pumain:2015}. However, the dominance tree
is constructed in order to highlight the spatial organization of urban systems
at various scales and while most methods used to determine relevant spatial
boundaries for cities rely on arbitrary thresholds that are different from one
country to another (shares of commuting flows, minimal distance between
consecutive built-up areas, etc.), this non-parametric method could provide an
alternative way to construct urban agglomerations with Voronoi cells at
different heights, and could constitute an interesting direction for future
studies.

\medskip

\paragraph*{Data and source code availability.} The data that support the
findings of this study are available for download at
\url{https://github.com/cestastanford/historical-us-city-populations} for the US
data, and at \url{https://www.insee.fr/fr/information/2414405} for the French
data. The source code is available at
\url{https://gitlab.huma-num.fr/tlouail/voronoize}.




\bibliographystyle{prsty}

\begin{thebibliography}{99}

\bibitem{Helsley:1990} Helsley, R. W., \& Strange, W. C. (1990). Matching and
  agglomeration economies in a system of cities. \emph{Regional Science and
    urban economics}, 20(2), 189-212.

\bibitem{Sanders:1997} Sanders, L., Pumain, D., Mathian, H., Guérin-Pace, F., \&
  Bura, S. (1997). SIMPOP: a multiagent system for the study of
  urbanism. \emph{Environment and Planning B: Planning and design}, 24(2),
  287-305.

\bibitem{Rozenblat:2007} Rozenblat, C., \& Pumain, D. (2007). Firm linkages,
  innovation and the evolution of urban systems. Cities in globalization:
  Practices, policies, theories, 130-156.
  
\bibitem{Zipf:1949} Zipf, G. K. (2016). Human behavior and the principle of
  least effort: An introduction to human ecology. Ravenio Books.

\bibitem{Gabaix:1999} Gabaix, X. (1999) Zipf's law for cities: an explanation. {\it The
    Quarterly journal of economics} {\bf 114}, 739-767.

\bibitem{Verbavatz:2020} Verbavatz, V. \& Barthelemy, M. (2020) The growth
  equation of cities. \emph{Nature} 587(7834): 397-401.

\bibitem{Cottineau:2017} Cottineau, C. (2017). MetaZipf. A dynamic meta-analysis
  of city size distributions. \emph{PloS one}, 12(8),
  doi.org/10.1371/journal.pone.0183919

\bibitem{Christaller:1933} Christaller, W. (1933) Die zentralen Orte in
  Suddeutschland: Eine okonomisch-geographische Untersuchung uber die
  Gesetzmafligkeit der Verbreitung der Siedlungen mit stddtischen Funktionen PhD
  thesis, Erlangen (Gustav Fischer, Jena)

\bibitem{Ullman:1941} Ullman, E. (1941). A theory of location for
  cities. \emph{American Journal of sociology}, 46(6), 853-864.


\bibitem{Haggett:1965} Haggett, P. (1965) Locational analysis in human
  geography. Edward Arnold Publishers, London.

\bibitem{Gonzalez-Val:2019} González-Val, R. (2019) The Spatial Distribution
  of US Cities. \emph{Cities}~\textbf{91}:
  157‑64. https://doi.org/10.1016/j.cities.2018.11.015.
  
\bibitem{Illian:2008} Illian, J., Penttinen, A., Stoyan, H., \& Stoyan,
  D. (2008) Statistical analysis and modelling of spatial point patterns. John
  Wiley \& Sons.

\bibitem{Jacobsen:2006} Jacobsen, M. (2006) Point process theory and
  applications: marked point and piecewise deterministic processes. Springer
  Science \& Business Media.

\bibitem{Moran:1950} Moran, P. A. (1950) Notes on continuous stochastic
  phenomena. \emph{Biometrika}, 37(1/2), 17-23.


 \bibitem{Haenggi:2012} Haenggi, M. (2012) Stochastic geometry for wireless
  networks. Cambridge University Press.

\bibitem{Okabe:2000} Okabe, A., Boots, B., Sugihara, K., \& Chiu, S. N. (2000)
  Spatial Tessellations: Concepts and Applications of Voronoi Diagrams (Second
  Edition). John Wiley \& Sons.

\bibitem{Laporte:book} Laporte G, Nickel S, Saldanha da Gama F (eds.)
  (2015). Location Science, Springer.

\bibitem{Bassolas:2020} Bassolas, A \& Nicosia, V. (2020) First-passage times to
  quantify and compare structural correlations and heterogeneity in complex
  systems. arXiv preprint arXiv:2011.06526.
  
\bibitem{Glass:1971} Glass, L., \& Tobler, W. R. (1971) Uniform distribution of
  objects in a homogeneous field: Cities on a plain.  \emph{Nature}, 233(5314),
  67-68 (1971).

\bibitem{Hernando:2014}
  Hernando, A. \& Plastino, A. (2014) Space–time correlations in urban sprawl.
  \emph{Journal of The Royal Society Interface}~\textbf{11}: 20130930.

\bibitem{Okabe:1996} Okabe A. \& Sadahiro Y. (1996) An illusion of spatial
  hierarchy: spatial hierarchy in a random configuration. \emph{Environment and
    Planning A}~28(9):1533--52.
  
\bibitem{Kadanoff:2000} Kadanoff, Leo P. (2000) Statistical physics: statics,
  dynamics and renormalization.  World Scientific Publishing Company.

\bibitem{USdata} U.S. Census Bureau and Erik Steiner, Spatial History
  Project. United States Historical City Populations, 1790-2010, Center for
  Spatial and Textual Analysis, Stanford University. (Data available at:
  \url{https://github.com/cestastanford/historical-us-city-populations},
  accessed on June 30,2021).

\bibitem{FRdata} Historique des populations communales. Recensements de la
  population 1876-2018~\url{https://www.insee.fr/fr/information/2414405}
  (accessed June 30, 2021).
  
\bibitem{Pumain:2015} Pumain D. et al (2015) Multilevel comparison of large
  urban systems. \emph{Cybergeo}, 706.  doi.org/10.4000/cybergeo.26730.

\bibitem{Cottineau:2016} Cottineau C., Hatna E., Arcaute E. \& Batty M. (2017)
  Diverse cities or the systematic paradox of Urban Scaling
  Laws. \emph{Computers, Environment and Urban Systems}~\textbf{63},
  80--94. doi.org/10.1016/j.compenvurbsys.2016.04.006.

\bibitem{Batty:2006} Batty, M. (2006) Rank clocks. \emph{Nature}~444:592-6.

\bibitem{Bretagnolle:2009} Bretagnolle A., Pumain D. \& Vacchiani-Marcuzzo
  C. (2009) The Organization of Urban Systems. In \emph{Complexity Perspectives
    in Innovation and Social Change}, 197--220,
  doi.org/10.1007/978-1-4020-9663-1\_7.

\bibitem{Gini} Dixon P.M., Weiner J., Mitchell-Olds T. \& Woodley R. (1987)
  \emph{Ecology}~\textbf{68}, 1548.

\bibitem{Abdi:2007} Abdi, H. (2007). The Kendall rank correlation
  coefficient. Encyclopedia of Measurement and Statistics. Sage, Thousand Oaks,
  CA, 508--510.
  
\bibitem{Chiu:1995} Chiu, S.N. (1995) Aboav-Weaire's and Lewis' laws—A review.
  \emph{Materials characterization} Mar 1;34(2):149-65.

\bibitem{Weaire:1984} Weaire, D. \& Rivier, N. (1984) Soap, cells and statistics
  --- random patterns in two dimensions. \emph{Contemporary Physics}, 25.1:
  59--99.

\bibitem{Pebesma:2018}Pebesma E. (2018) Simple Features for R: Standardized Support for
  Spatial Vector Data. \emph{The R Journal}, 10(1), 439–446. doi:
  10.32614/RJ-2018-009.

\bibitem{Wickham:2016} Wickham H. (2016) ggplot2: Elegant Graphics for Data
  Analysis. Springer-Verlag New York. ISBN 978-3-319-24277-4,
  https://ggplot2.tidyverse.org.

\bibitem{Strano:2020} Strano E., Simini F., De Nadai M., Esch T. and Marconcini
  M. (2020) The agglomeration and dispersion dichotomy of human settlements on
  Earth. arXiv:2006.06584.
  
\bibitem{Xiao:2022} Xiao Y. and Gong P. (2022) Removing spatial autocorrelation
  in urban scaling analysis. \emph{Cities},
  124. https://doi.org/10.1016/j.cities.2022.103600
  
   
   


  
\end{thebibliography}


\section*{Author Contributions}
TL and MB designed the study. TL wrote the code and performed calculations. TL
and MB analyzed and interpreted the results, prepared the figures and wrote the
manuscript.

\section*{Competing Interests}
 The authors declare that they have no competing interests.

\section*{Acknowledgements}

The authors thank the anonymous reviewers for their useful comments that helped
improving the manuscript. TL thanks the many developers of the free software
packages that were used in this study, especially \emph{sf}~\cite{Pebesma:2018},
GDAL, GEOS and PROJ to perform the geocomputation and
\emph{ggplot2}~\cite{Wickham:2016} to draw the figures. Last but not least, TL
warmly thanks the Ramelet family for hosting him in the winter of 2021 during
the COVID-19 lockdown, during which most of this work was done.
 

\end{document}